\documentclass[11pt,twoside]{article}
\usepackage{atmp}
\copyrightnotice{2003}{7}{727}{749}	
\newcommand{\bj}{{\bar{j}}}
\newcommand{\bi}{{\bar{i}}}
\newcommand{\tpsi}{{\tilde{\psi}}}

\newcommand{\al}{{\alpha}}
\newcommand{\la}{{\lambda}}
\renewcommand{\d}{{\rm d}}
\newcommand{\ep}{{\epsilon}}
\newcommand{\bep}{{\bar\ep}}

\newcommand{\Ga}{\Gamma}

\def\deg{{\rm deg}}

\renewcommand{\l}{\left}
\renewcommand{\r}{\right}

\def\part{\partial}

\newcommand{\Tr}{{\rm Tr}}
\newcommand{\str}{{\rm STr}}
\renewcommand{\th}{{\theta}}
\newcommand{\bz}{{\bar{z}}}

\newcommand{\CC}{{\mathbb C}}

\newcommand{\ZZ}{{\mathbb Z}}

\newcommand{\Hom}{{\rm Hom}}
\newcommand{\Mor}{{\rm Mor}}
\newcommand{\End}{{\rm End}}
\newcommand{\Ext}{{\rm Ext}}
\newcommand{\cS}{{\mathcal S}}
\newcommand{\cC}{{\mathcal C}}

\newcommand{\tA}{{\tilde A}}
\renewcommand{\L}{{\mathcal L}}
\newcommand{\N}{{\mathcal N}}

\newcommand{\D}{{\mathcal D}}
\newcommand{\T}{{\mathcal T}}
\renewcommand{\O}{{\mathcal O}}

\newcommand{\A}{{\mathcal A}}
\newcommand{\ra}{\rightarrow}
\newcommand{\bpartial}{{\bar\partial}}
\newcommand{\no}{\nonumber}
\newcommand{\op}{\oplus}
\newcommand{\ot}{\otimes}
\newcommand{\perf}{{\mathsf{Perf}}}
\newcommand{\id}{{\rm id}}
\newcommand{\FL}{{(-1)^{F_L}}}
\newcommand{\dlr}{{\stackrel{\leftrightarrow}{\partial}}}

\begin{document}
 \setcounter{page}{727}
\title{Topological Correlators
in Landau-Ginzburg Models with Boundaries}

\url{hep-th/0305136}	
 
\author{Anton Kapustin$^1$ and Yi Li$^2$}		
\address{$^1$California Institute of Technology\\
	Department of Physics\\Pasadena, CA 91125}
\addressemail{kapustin@theory.caltech.edu}	
	
\address{$^2$California Institute of Technology\\
	Department of Physics
	\\Pasadena, CA 91125}
\addressemail{yili@theory.caltech.edu}

\markboth{\it Topological Correlators in LG Models with Boundaries\ldots}
		{\it A. Kapustin and Y. Li}

\begin{abstract}
We compute topological correlators in Landau-Ginz\-burg models on a Riemann surface
with arbitrary number of handles and boundaries. The boundaries may correspond to arbitrary
topological D-branes of type B. We also allow arbitrary operator insertions on the boundary and
in the bulk.
The answer is given by an explicit formula which can be regarded as an open-string generalization
of C.~Vafa's formula for closed-string topological correlators.
We discuss how to extend our results to the case of Landau-Ginzburg orbifolds.
\end{abstract}
 
\cutpage

\section{Introduction}

Topological Landau-Ginzburg models are an important class of 2d topological field theories (2d TFTs).
When coupled to topological gravity, they give soluble examples of noncritical string backgrounds.
They are also closely related to topological sigma-models describing superstring propagation
in Calabi-Yau manifolds (the so-called CY/LG correspondence). 

C.~Vafa showed how to compute topological correlators in Landau-Ginzburg (LG) models on Riemann surfaces of 
arbitrary genus~\cite{Vafa}. Remarkably, the answer is given by a simple closed formula. The goal
of this paper is to generalize the results of Ref.~\cite{Vafa} to Riemann surfaces with boundaries.
Topological boundary conditions for a 2d TFT are also called topological D-branes (of type B, in this case).
In the case of LG models, they have been studied in Refs.~\cite{GSS,HIV,Hbook,Hlinear,BH}.
In order to get a simple formula for topological correlators, one first needs a simple description
of topological D-branes. Recently such a description was proposed by M.~Kontsevich (unpublished). 
According to this proposal, topological D-branes are CDG modules over a certain CDG algebra which
depends on the superpotential $W$ of the Landau-Ginzburg model. Spaces of open strings stretched between
D-branes are identified with the space of morphisms in the derived category of CDG modules.\footnote{Note
for physicists: do not panic! It is much simpler than it sounds.} The OPE algebra of boundary 
operators is encoded in the composition law of morphisms. It should be noted that the language of
derived categories is very suited to the discussion of topological D-branes and has been used in the
context of open string field theory (see e.g. Refs.~\cite{LazBK,LazU}).

In our previous paper~\cite{us}, we have given a physical derivation of Kontsevich's conjecture. We also performed
some explicit computations in the case when the LG model is massive. The derivation is reviewed in 
Section~\ref{CDG} below, with some improvements in exposition. The main physical idea is to represent 
topological D-branes as bound states
of branes and anti-branes of maximal dimension. Kontsevich's conjecture concerning the boundary OPE
follows from the analysis of the world-sheet action for the LG model with boundaries in the presence of 
both branes and anti-branes. 

>From the physical viewpoint, computing the boundary OPE is only part of the problem. One would also
like to know the topological metric on the space of open strings. Together with the boundary
OPE, this provides complete information about topological boundary correlators on a disk.
More generally, one may consider disk correlators involving both bulk and boundary insertions. Even more
generally, one may consider correlators on a Riemann surface with arbitrary number of boundaries
and handles, and arbitrary insertions of bulk and boundary operators. In this paper we give explicit 
formulas for all these correlators. In some special cases, topological disk correlators have been computed
in Ref.~\cite{Hbook}.

>From the mathematical viewpoint, an interesting result of this paper is a simple explicit formula for the
topological open-string metric, i.e. for a non-degenerate pairing between vector spaces $\Mor(A,B)$ and $\Mor(B,A)$,
where $A$ and $B$ are arbitrary B-branes (i.e. CDG modules), and $\Mor(A,B)$ is the space of morphisms from $A$ to $B$
in the derived category
of CDG modules. The non-degeneracy of the pairing follows from the properties of the path integral defining
topological LG models, or equivalently, from the axioms of 2d TFT with boundaries~\cite{MS,Laz}. The pairing
is natural, in an obvious sense. In particular, the algebra of
endomorphisms of any B-brane is naturally a graded Frobenius algebra. Unlike in the closed-string case, this
algebra is not supercommutative, in general.

To appreciate better the significance of the open-string metric, recall that in the case of B-branes on a Calabi-Yau
manifold $X$ this metric is given by the usual pairing between $\Ext^k(A,B)$ and $\Ext^{n-k}(B,A)$, where $A$ and $B$
are objects of the bounded derived category of $X$. The pairing
is non-degenerate as a consequence of the Serre duality theorem. On a general compact complex manifold Serre
duality tells us that there is a non-degenerate pairing between $\Ext^k(A,B)$ and $\Ext^{n-k}(B,A\ot K)$, where
$K$ is the canonical line bundle. Guided by this example, the authors of Ref.~\cite{BK} introduced the notion of
a Serre functor on a triangulated category. Basically, a Serre functor $\cS$ on a triangulated category $\cC$ is an 
autoequivalence of $\cC$ such that there exists a natural non-degenerate pairing between $\Mor(A,B)$ and $\Mor(B,\cS(A))$.
A Serre functor on $\cC$, if it exists, is unique. Using this terminology,
one can say that all categories of topological branes have a trivial Serre functor. This is interesting, because
such categories are hard to come by. For example, the category of B-branes in a Landau-Ginzburg model must have
a trivial Serre functor, and our result for the open-string metric gives an explicit formula for the corresponding
pairing. The non-degeneracy of this pairing is far from obvious, and it would be interesting to find a direct
mathematical proof of this fact.\footnote{Topological metric for closed string states has been computed by C.~Vafa~\cite{Vafa}.
Its non-degeneracy follows from the local duality theorem~\cite{GH}.}

\section{B-branes in LG models as CDG modules}

\subsection{Physics}

In this section we review the physical derivation of the boundary OPE algebra in the topologically twisted LG model. Let $X\simeq \CC^n$, and 
$W$ be a holomorphic function on $X$ (the superpotential). The worldsheet bulk Lagrangian density is given by (cf. \cite{HIV})
\begin{eqnarray}
	\L &=& \big|\part_0\phi^i\big|^2-\big|\part_1\phi^i\big|^2 + \frac{i}{2}\l(\psi_-^\bi\dlr_+\psi_-^i+\psi_+^\bi\dlr_-\psi_+^i \r)\no\\
	&& -\frac14\big|\part_i W\big|^2-\frac12(\partial_i\partial_jW)\psi_+^i\psi_-^j-\frac12(\partial_\bi\partial_\bj\bar{W})\psi_-^\bj\psi_+^\bi \no
\end{eqnarray}
For definiteness, we have chosen the standard flat Kahler metric on $X$. If the worldsheet $\Sigma$ has no boundary, the action is invariant 
up to a total derivative under the $(2,2)$ worldsheet supersymmetry. What will be important to us is the type-B $\N=2$ SUSY transformation
\begin{align}
&\delta\phi^i = \ep\psi^i, &\delta\phi^{\bi} = -\bep\psi^{\bi} & \no\\
&\delta\psi^i = -2i\bep\dot{\phi}^i, &\delta\psi^{\bi} = 2i\ep\dot{\phi}^{\bi} & \no
\end{align}
Our conventions are the same as in Ref.~\cite{HIV}.
The presence of worldsheet boundaries breaks $\N=2$ SUSY in general. For instance, a brane wrapped on the whole target space $X$ 
is well known to be non-supersymmetric. However it is not unreasonable to expect that all supersymmetric D-branes (in our case, B-branes) can be 
obtained by the process of tachyon condensation on top-dimensional brane-anti-brane systems. In fact this is the basic assumption that 
we will make in this paper. As we shall see, the requirement of type-B $\N=2$ SUSY invariance after tachyon condensation naturally gives rise to 
the mathematical structure of boundary OPE algebra mentioned above.

More specifically, we want to add a boundary term to the action for the tachyon and gauge fields associated with a brane-anti-brane system which 
wrap the whole space $X$, such that the type-B supersymmetry can be restored. To this end, we first modify the bulk Lagrangian density by a 
total derivative
$$\L\;\to\; \L+\frac{i}2\l(W-\bar{W}\r)'$$
This does not change the physics in the bulk, although it makes easier to write down the boundary action. Taking the standard Neumann 
boundary conditions for all bulk fields, the modified bulk action, $S$, has the following variation under type-B SUSY:
\begin{eqnarray}
	\delta S &=& \int_{\partial\Sigma}\,\frac{i}2\bep\l(\psi^{\bar{i}}\partial_{\bar{i}}{\bar{W}}+\psi^i\partial_i{W}\r) \;+\;{\rm h.c.}\no
\end{eqnarray}
On the other hand, the modified bulk action is invariant under an $\N=1$ subalgebra: $\ep=i\ep_1, \bep=-i\ep_1$. This means that the boundary term 
we should add to recover $\N=2$ invariance must be $\N=1$-supersymmetric by itself.

Let the brane-anti-brane system be described by a pair of Hermitian vector bundles $E=(E_1,E_2)$ with unitary connections. We may regard 
$E$ as a $\ZZ_2$-graded vector bundle. Since  $X\simeq \CC^n$, these bundles are necessarily trivial.
For simplicity, let us assume that $E_1$ and $E_2$ have rank one. 
The generalization to higher rank is discussed below. The boundary degrees of freedom are described by a boundary fermion $\eta$, 
which is taken to be a section of $\Hom(E_2,E_1)$. This field couples to a boundary tachyon $T$, which is a section of the bundle 
$\Hom(E_1,E_2)$, and the gauge field $A$ which is a connection on the line bundle $\Hom(E_2,E_1)$. The gauge field $A$ is
the difference of gauge fields on the brane and the anti-brane. The boundary action also depends on the sum of the two
gauge fields, which we denote $\tA$. The field $\tA$ is a connection on the line bundle $E_1\ot E_2$. The $\N=1$-supersymmetric boundary
action has the standard form \cite{ttu}
$$S_b = \frac12\int_{\partial\Sigma}\d x^0\d\th \l[\bar{\Ga}\l(D_\th - i A_I(\Phi)D_\th\Phi^I\r)\Ga + \Ga\T+\bar{\T}\bar{\Ga}+i\tA_I D_\th \Phi^I\r]$$
where $D_\th \;=\; {\part_\th} -2i \th{\part_0}$, and $\T$, $\Ga$ are superfield completions:
\begin{eqnarray}
\Ga &=& \eta+i\th P\no\\
\T &=& T+i\th\psi^{I}\part_{I}T\no
\end{eqnarray}
Expressing in terms of components after integrating out the massive field $P$, one gets 
\begin{equation}
S_b = \int_{\partial\Sigma}\l[i\bar{\eta}(\part_0-iA_0)\eta + \l(\frac{i}2\psi^I D_I T\eta+{\rm h.c.}\r) - \frac12 T\bar{T}+\tA_0\r]
\label{eq:Sb}
\end{equation}
where $D_I$ is the covariant derivative associated with the connection $A$, $A_0$ is given by 
$$A_0 = A_I\part_0\phi^I-\frac{i}4F_{IJ}\psi^I\psi^J,$$
and $\tA_0$ is defined similarly, but with $A_I$ replaced with $\tA_I$.

Now we require that the total action $S+S_b$ preserves the type-B $\N=2$ supersymmetry. The variations of $T$ and $A$ are determined by variations of 
the bulk fields $\phi^i$. The only field whose $\N=2$ variation is not known is the boundary fermion $\eta$. However we know that its 
variation cannot involve $\bar{\eta}$ since such a term destroys $\N=2$ invariance. Gauge invariance then dictates the following general form
\begin{equation}
\delta\eta \;=\; i\al\eta + \ep s_1+\bep s_2
\label{eq:eta}
\end{equation}
where $\al$ is an unknown function, and $s_{1,2}$ are a pair of sections on $\Hom(E_2,E_1)$ which do not depend on $\eta$. It is not difficult 
to see that $\N=2$ invariance implies $\delta A_0 = \part_0\al$, which in turn is satisfied if and only if
$$F_{ij}=0=F_{\bi\bj}, \qquad \al = \ep A_i\psi^i-\bep A_{\bi}\psi^{\bi}.$$
In other words, $\N=2$ invariance requires that the connection $A$ defines a holomorphic structure on the line bundle $\Hom(E_2,E_1)$.
Similarly, $\N=2$ invariance requires that the curvature of $\tA$ be of type $(1,1)$. Thus both $E_1$ and $E_2$ must be holomorphic
line bundles.

To have $\N=2$ invariance one must further demand that all $\eta$-dependent (resp. $\bar\eta$-dependent) terms in $\delta S_b$ cancel each other. 
It can be shown that this amounts to two conditions on $s_1$ and $s_2$. First, they must be purely bosonic, i.e. they depend only on $\phi$ and 
not on $\psi$. Second, they must satisfy the following equations 
\begin{eqnarray}
D_0\bar{s}_1 &=& -i\dot{\phi}^iD_iT\no\\
D_0\bar{s}_2 &=& i\dot{\phi}^{\bi}D_{\bi}T\no
\end{eqnarray}
These two equations have covariant solutions if and only if $T$ takes the following form 
$$T = F+\bar{G}, \qquad D_{\bi}F\;=\;0, \quad D_{i}\bar{G}\;=\;0$$
together with the identification
$$s_1=i\bar{F}, \qquad s_2=-iG.$$
One can also show that these conditions are sufficient for cancellation of $\eta$-dependent and $\bar{\eta}$-dependent terms in $\delta S_b$.

The final requirement of $\,\N=2$ invariance comes from demanding that the remaining terms in $\delta S_b$, which now do not involve $\eta$ at all, 
exactly cancel the bulk variation $\delta S$. It is straightforward to show that the variation of $S_b$ is given by
$$\delta S_b = \int_{\partial\Sigma} \frac12\bep\l(\psi^{\bar{i}}\partial_{\bar{i}}(\bar{F}\bar{G}) -\psi^i\partial_i(FG) \r) + {\rm h.c.}$$
which cancels the bulk variation if and only if 
$$FG \;=\; iW + {\rm const.}$$

If the bundles $E_1$ and $E_2$ have rank higher than one, there is an additional complication. Namely, the integrand in the string path integral
is no longer an exponential of a local action: there is also a non-local factor
$$
\Tr P\exp \left(\oint \left(A_\mu \frac{d\phi^\mu}{d\tau}+\dots\right) d\tau \right),
$$
where dots denote terms depending on the fermions and the boundary tachyon. In order to figure out the conditions imposed by $\N=2$ SUSY, it is desirable
to have a local action. This can be achieved by representing the path-ordered exponential by a path-integral over auxiliary
degrees of freedom living on the boundary of the worldsheet. In fact, there are several way to do this; for example, the
boundary degrees of freedom can be bosonic or fermionic. The action for all these fields,
and their transformation properties under $\N=2$ SUSY will be described elsewhere. The results are the following. First of all,
both $E_1$ and $E_2$ must be holomorphic vector bundles. Second,
the boundary tachyon, which is still a section of $\Hom(E_1,E_2)$, must have the form
$$
T=F+G^\dag,
$$
where $F$ is a holomorphic section of $\Hom(E_1,E_2)$, and $G$ is a holomorphic section of $\Hom(E_2,E_1)$.
Finally, $F$ and $G$ must satisfy
$$
FG=iW\cdot\id_{E_2}+{\rm const},\quad GF=iW\cdot \id_{E_1}+{\rm const}.
$$
This implies, among other things, that $E_1$ and $E_2$ must have the same rank.

In the B-twisted model, the anti-holomorphic fermions $\psi^\bi_+$ and $\psi^\bi_-$ become 0-forms on the world-sheet, while the holomorphic 
components $\psi^i_+$ and $\psi^i_-$ combine into a 1-form $\rho^i$. The BRST transformation is obtained from type-B $\,\N=2$ variation by 
setting $\ep=0$. The BRST charge $Q$ is the sum of the bulk and boundary contributions. The bulk contribution is standard. The boundary part 
$Q_{\rm boundary}$ must generate the correct transformation 
$$
\delta\eta \;=\; -i\bep A_\bi\psi^\bi \,-\,i\bep G
$$
Because the connection defines a holomorphic structure on $\Hom(E_1,E_2)$, one can locally trivialize the connection to $A_\bi = 0$. Since 
$X$ is affine, one can actually take $A_\bi=0$ globally. The boundary BRST charge in this gauge is given by
$$Q_{\rm boundary} \;=\; -iF\eta \,+\, iG\bar{\eta}$$
Since $A_\bi=0$, $F$ and $G$ are simply holomorphic matrix-valued functions of $\phi^i$. 

One may check explicitly that the BRST invariance is restored after adding the boundary contribution. The bulk BRST charge is known to be given by
$$Q_{\rm bulk} = \int dx^1 \l(\psi^\bi\part_0{\phi}^i - \tpsi^\bi\part_1\phi^i + \frac{i}2 \tpsi^i\part_iW\r)$$
where $\tpsi\equiv \psi_--\psi_+$. The bulk charge is not nilpotent but satisfies
$$Q_{\rm bulk}^2 \;=\; -iW|_{\part\Sigma}$$
The total BRST charge squares to zero by virtue of $Q_{\rm boundary}^2=iW$, as well as $\{Q_{\rm bulk},Q_{\rm boundary}\}=0$, 
which comes from 
the fact that $F$ and $G$ are holomorphic.

Physical states in the topological theory must be annihilated by the total BRST
charge. Since we are working with a topological theory, the zero-mode approximation is adequate. The zero-mode Hilbert space 
consists of sections 
of $\End(E)$. The operators $\eta$ and $\bar\eta$ are realized by the block matrices
$$
\eta=\begin{pmatrix}  0 & i\\ 0 & 0 \end{pmatrix}, \quad 
\bar\eta=\begin{pmatrix}  0 & 0\\ -i & 0 \end{pmatrix}
$$
The boundary part of the BRST operator is therefore identified with
\begin{equation}
D\;=\,\begin{pmatrix}  0 & F\\ G & 0 \end{pmatrix}
\label{eq:DFG}
\end{equation}
while the bulk part reduces to the standard Dolbeault operator
$$
Q_{\rm bulk} \;=\; \bpartial \;\equiv\; d\phi^{\bi}\part_{\bi}.
$$
A section $\phi$ of $\End(E)$ is BRST-invariant\footnote{Do not confuse the section $\phi\in\End(E)$ with the bulk fields $\phi^i$.} if and 
only if $\phi$ is holomorphic
and super-commutes with $Q_{\rm boundary}$. If $\phi$ is a super-commutator
of $Q_{\rm boundary}$ and a holomorphic section $\alpha$, then it is BRST trivial. Let us denote by $V$ the space of holomorphic sections of 
$\End(E)$. Since we are working on $\CC^n$, this is simply the space of $2\times 2$ block matrices with holomorphic entries. It has a natural 
decomposition into even and odd (diagonal and off-diagonal) components. Let $\D$ denote the following operator on $V$:
$$
\D: \; V \mapsto [Q_{\rm boundary},V]
$$
where the brackets denote the super-commutator. It squares to zero by virtue of $FG=GF=W.$\footnote{We absorb the innocuous factor of $i$ in $W$ 
in what follows.} In view of the above, the space of physical states can be identified with the cohomology of $\D$. Obviously, the OPE algebra
corresponds simply to matrix multiplication of elements of $V$.

\subsection{Math}
\label{CDG}

Now let us reformulate the structure which appeared above in mathematical terms.
A CDG algebra is a triple $(A,d,B)$, where $A$ is a $\ZZ_2$-graded associative algebra,
$d$ is an odd derivation of $A$, and $B$ is an even element of $A$ satisfying $d^2a=[B,a]$
for any $a\in A$. Here the brackets denote the super-commutator. A CDG-module over a CDG algebra
is a pair $(M,D)$, where $M$ is a $\ZZ_2$-graded module over $A$, and $D$ is an odd endomorphism
of the $\ZZ_2$-graded vector space $M$ which satisfies the super-Leibniz identity
$$
D(a\cdot m)=da\cdot m-(-1)^{|a|}a\cdot Dm,\quad \forall a\in A, \,m\in M,
$$\
and the ``twisted differential'' condition
$$
D^2 m = B\cdot m, \quad \forall m\in M.
$$
In the special case $B=0$ CDG algebras and modules are called DG algebras and modules.

In the above definition one can replace $\ZZ_2$-grading with $\ZZ$-grading. Then one requires
$d$ and $D$ to be of degree one, and $B$ must be of degree two. In what follows, we
will be working with $\ZZ_2$ gradings, unless specified otherwise. 

A well-known example of a ($\ZZ$-graded) DG algebra is the algebra of differential forms on a smooth manifold,
equipped with the de Rham differential. More generally, one may consider the algebra of sections of the
endomorphism bundle of a flat vector bundle with a connection, equipped with the flat covariant differential
(the so-called twisted de Rham complex). CDG algebras appear when we replace flat connections
with arbitrary connections. In this example $B$ is simply the curvature 2-form of the connection.
This is the origin of the term ``CDG algebra'' (CDG means ``curved differential graded''). 

Let us make three trivial remarks about the definitions of CDG algebras and modules. First, if $B\neq 0$,
then a CDG algebra cannot be naturally regarded as a CDG-module over itself, since the equation $d^2a=[B,a]$ is in conflict
with the desired $D^2a=Ba$. Second, if the algebra $A$ is super-commutative, then the pair $(A,d)$ is
a DG-algebra, but DG modules over this DG-algebra are not the same as CDG-modules over the CDG-algebra $(A,d,B)$,
unless $B=0$. Indeed, in the former case we have the condition $D^2m=0$, while in the latter case we have
$D^2m=Bm$. Third, if $d=0$, but $B\neq 0$, then $D$ is simply an odd endomorphism of the module $M$
(i.e. $D(a\cdot m)=(-1)^{|a|} a\cdot D(m)$ for all $a$ and $m$) satisfying $D^2=B$.

The results of the previous section can be rephrased as follows. Consider a CDG algebra $\A=(\O,0,W)$,
where $\O$ is the algebra of polynomials on $\CC^n$, and $W$ is the superpotential. The graded bundle
$M=E=E_+\op E_-$ which describes the brane-anti-brane system is a graded module over $\O$, and the
tachyon field $D\in \End(E)$ gives it the structure of a CDG module over $\A$. Since we are working
on $\CC^n$, the module $M$ is simply a sum of several copies of $\O$ (i.e. a free module), and $D$ can be viewed
as a matrix with elements in $\O$.

In order to rephrase in this language the boundary OPE algebra, we need to discuss endomorphisms of CDG
modules. We are interested in the situation when $A$ is purely even and commutative, and $d=0$. 
Consider first endomorphisms of $M$ regarded simply as an $A$-module. These are linear maps from $M$ to $M$
which commute with the action of $A$. If $M$ is a free module of rank $r$ (i.e. a sum of $r$ copies of $A$
regarded as a module over itself), $\End_A(M)$ is the algebra of $r\times r$ matrices with entries in $A$.
This algebra has natural $\ZZ_2$-grading: endomorphisms which do not change the grading on $M$ are declared 
even, while those which change the grading are declared odd. In other words, if $M=M_+\op M_-$ where
$M_+$ and $M_-$ are free modules of rank $r_+$ and $r_-$, $r_++r_-=r$, then endomorphisms which map
$M_+$ to $M_+$ and $M_-$ to $M_-$ are even, while those which map $M_+$ to $M_-$ and vice versa are odd.
In the matrix notation, we can write an endomorphism $\phi$ as a block matrix
$$
\phi=\begin{pmatrix}  A & B\\ C & D  \end{pmatrix},
$$
where $A,B,C,D$ are matrices with entries in $A$ of sizes $r_+\times r_+$, $r_+\times r_-$, $r_-\times r_+$,
and $r_-\times r_-$, respectively. With the above grading, $A$ and $D$ are even, and $B$ and $C$ are odd.

The endomorphism algebra has a natural odd derivation which squares to zero. It is given by
$$
\D: \phi\mapsto [D,\phi],
$$
where the brackets denote the supercommutator. Note that $\D^2=0$ because of the super-Jacobi identity and the fact that
$D^2=W.$ To summarize, if $M$ is a CDG module, then $\End_A(M)$ is a DG algebra. 

It is easy to see that in the case $A=\O$, $B=W$, this DG algebra is precisely the algebra of boundary operators 
in the zero-mode approximation.
The differential $\D$ is identified with the boundary part of the BRST operator. Thus the space of physical
states is simply the cohomology of $\D$, and the boundary OPE structure is simply the algebra structure on the cohomology
of $\D$. 

One can easily generalize this to the case when strings begin on one brane and end on another one. In this case
we have a pair of CDG modules $M$ and $N$, and the space of morphisms between them is a graded vector space $\Hom_A(M,N)$
equipped with a differential $\D$ defined as follows:
$$
\D:\phi\mapsto D_N\circ\phi - (-1)^{|\phi|}\phi\circ D_M, \quad \phi\in \Hom_A(M,N).
$$
This vector space is identified with the space of operators which change the boundary conditions
(in the zero-mode approximation). The cohomology corresponding to the differential $\D$ is identified with the
BRST cohomology (i.e. the space of physical boundary changing operators modulo BRST trivial ones). If we have three
different branes, say, $M,N$, and $K$, we can compose morphisms from $M$ to $N$ and from $N$ to $K$ to obtain morphisms
from $M$ to $K$. This composition descends to the cohomology of $\D$, as one can easily see, and corresponds to composition
of boundary-changing operators.

Consider now a category whose objects are B-branes, whose spaces of morphisms are spaces of physical states
for strings stretched between branes, and where composition of morphisms is defined by means of product
of boundary-changing operators. Consider also a category whose objects are free CDG modules over $A=\O$,
whose morphisms are cohomology classes of the differential $\D$, and where composition is the obvious one.
The above discussion can be restated by saying that these two categories are equivalent. The second category
coincides with the derived category of CDG modules over the CDG algebra $(\O,0,W)$. Thus we have derived 
Kontsevich's conjecture.
The main physical assumption in the derivation is that any B-brane can be obtained as a bound state of branes
and anti-branes of maximal dimension (D-branes wrapping the whole target space). 

>From the mathematical
point of view, there is the following subtlety: the derived category is not simply a category, but has some
extra structure (the shift functor and the so-called distinguished triangles). This structure makes
the derived category into a triangulated category. Ideally, one would like to show that triangulated structures
also agree. One can easily see that the shift functor on the derived category corresponds to the functor
on the B-brane category which exchanges branes and anti-branes, i.e. exchanges $F$ and $G$ in Eq.~(\ref{eq:DFG}).
As for distinguished triangles, they should show up in physics if we consider stability conditions
for B-branes~\cite{Douglas}. We hope to return to this issue in the future.

\subsection{Aftermath}

As discussed in Ref.~\cite{us}, the category of B-branes is non-zero only if the set $W=W_0$
is the critical level set of $W$, i.e. it contains some critical points of $W$. Furthermore,
it is sensitive only to the infinitesimal neighborhood of the critical points. It turns out
one can make a very precise mathematical statement about how this localization occurs, i.e. one can
describe the category of B-branes entirely in terms of the geometry of the critical level set~\cite{orlov}.
Unfortunately, the mathematical statement is not so easy to understand for a non-mathematician.
Nevertheless, we will try to give some idea about it, because in some cases it can be used very effectively
to determine the structure of the category of B-branes (see below).

Consider the critical level $W=W_0$. We stress that it should be regarded as an algebraic variety,\footnote{The more precise word here is 
a {\em scheme}, since the structure sheaf of the critical level $\{W=W_0\}$ contains nilpotent elements. However we shall use the less accurate but hopefully more familiar 
term {\em variety} in this paper to avoid scaring away the readers. Here we make the standard (but often unjustified) assumption
that someone is still reading the paper.} rather
than as a set. For example, if $W=z^n$, the set defined by $W=0$ consists of a single point, and
does not depend on $n$. On the other hand, the algebra of functions on the algebraic variety
${\rm Spec}(\CC[z]/W)$ is $\CC[z]/z^n$, and depends on $n$.

If $W_0$ is a critical value of $W$, the variety $Z$ given by $W=W_0$ is singular. In Ref.~\cite{orlov} the following
construction was proposed. Consider the bounded derived category of coherent sheaves on $Z$.
Its objects are bounded complexes of coherent sheaves on $Z$. A perfect complex is an object of $D^b(Coh(Z))$
which is quasi-isomorphic to a bounded complex of locally free sheaves. One may consider a full subcategory
$\perf$ of $D^b(Coh(Z))$ whose objects are perfect complexes, and also the quotient category $D_{Sg}(Z)=D^b(Coh)Z))/\perf$.
The point of this construction is that for a smooth $Z$ the category $\perf$ coincides with the whole
$D^b(Coh(Z))$, and $D_{Sg}(Z)$ is trivial. Thus $D_{Sg}(Z)$ is only sensitive to singularities of $Z$.

The main result of Ref.~\cite{orlov} is that the category of B-branes is equivalent to $D_{Sg}(Z)$
where $Z$ is the critical variety corresponding to $W$. This result gives a precise meaning to the
physical statement that strings are confined to the critical points of $W$. 

In particular, in the case $W=z^n$, corresponding to the A-series of $N=2$ minimal models, the variety $Z$ is
zero-dimensional, and one can completely describe the category $D_{Sg}(Z)$. As shown in Ref.~\cite{orlov},
the category of B-branes is semi-simple, i.e. every object is a finite direct sum of indecomposable objects.
Furthermore, there are $n-1$ indecomposable B-branes labeled by $k=1,\ldots,n-1$, and the endomorphism
algebra of the $k$-th brane (i.e. the algebra of boundary operators which do not change the boundary condition)
is $\CC[z,\theta]/(z^p,\theta^2+z^{n-2p})$, where $z$ is an even (bosonic) variable, $\theta$ is an odd (fermionic)
variable, and $p=min(k,n-k)$. We discuss this and other examples in detail elsewhere.
Here we just remark that although it is easy to construct all irreducible B-branes as
CDG modules, it is far from obvious that {\it any} B-brane is isomorphic to a sum of these. On the other hand,
the results of Ref.~\cite{orlov} quickly reduce the problem of classification of branes to the problem
of classification of matrix solutions to the equation $P^n=0$. The latter is solved by the Jordan normal
form theorem.

\section{Topological correlators in LG models with boundaries}

\subsection{Disk correlators}
In this section we derive a closed-form formula for disk correlators of 
arbitrary bulk and boundary topological operators. The crucial element of the derivation is to represent the B-brane on which 
the boundary of the disk ends as a brane-anti-brane system with tachyon fields turned on, as discussed in Section 2. 
For simplicity, we assume the target space $X$ to be affine. The brane-anti-brane pair is denoted by $E=(E_1,E_2)$ with 
$E_1,E_2$ being equal rank holomorphic vector bundles. The formula we seek to prove is
\begin{equation}
	\big\langle\alpha\cdot\phi\big\rangle_{\rm disk} \;=\;\frac1{n!(2\pi i)^n} \oint \frac{\alpha \cdot 
\str\big[(\partial D)^{\wedge n}\phi\big]}{\part_1W\part_2W\ldots\part_nW}
\label{eq:form}
\end{equation}
A few comments are in order. The bosonic bulk chiral fields are renamed here by $z_1, z_2,\ldots,z_n$, and $\alpha$ is an arbitrary bulk operator represented by an element of $\CC[z_1,\ldots,z_n]/I$, with $I$ being the ideal 
generated by all the differentials $\part_i W$. The twisted differential $D$ is defined as in (\ref{eq:DFG}), and 
$\phi\in\End(E)$ is an arbitrary boundary operator, i.e. a representative of a class in the $\D$-cohomology. The supertrace ``$\str$'' is 
a twisted trace operator in the fundamental representation of $E_1\oplus E_2$:
$$\str\,\phi = {\rm Tr}\;(-1)^F\phi$$
where $(-1)^F\in\End(E_1\oplus E_2)$ is the grading operator defined by
$$(e_1,e_2)\mapsto(e_1,-e_2).$$
Thus the integrand in Eq.~(\ref{eq:form}) is a meromorphic $n$-form on $\CC^n$.
Finally, the contour integral is performed over a union of $n$-dimensional Lagrangian tori  
which enclose all the critical points of $W(z)$~\cite{Vafa,GH}. This integral can be regarded
as a generalized residue, and can also be written as an integral of a $2n-1$-form over a large $2n-1$-dimensional sphere in $\CC^n$.
The $2n-1$-form is the product of the numerator in Eq.~(\ref{eq:form}), which is a holomorphic $n$-form, and a certain $n-1$-form
depending on $W$ alone whose explicit form can be found in Ref~\cite{GH}. This way of writing the generalized residue emphasizes
that it can be computed using only the asymptotic behavior of $D,W$, and $\phi$. In particular, no explicit knowledge
of the critical points of $W$ is required. In the case $n=1$, the two ways of writing the formula become almost identical:
in both cases one has an integral of a meromorphic 1-form along a contour in the complex plane, and the only difference is
whether the contour consists of several small circles enclosing all the singularities of the 1-form, or is a single large circle
enclosing all the singularities. In the next section, we shall generalize (\ref{eq:form}) to Riemann surfaces with arbitrary number 
of handles and boundaries.

In order to evaluate the path integral on a disk, we will use the fact that it is a topological invariant and so does not depend on the 
worldsheet metric $g_{\alpha\beta}$. Therefore we may rescale the metric by $g\to\la^2g$ and evaluate the integral in the limit $\lambda\ra 0$. 
It is also well-known that the path integral is localized 
around instantons which are 
fixed points of the BRST operator \cite{Witten_Mir}. In our case these instanton configurations are given by
$$\part_iW=0, \qquad D=0.$$
Around each of these instantons, the contributions of bosonic and fermionic non-zero modes cancel each other, and the path integral 
is reduced to an ordinary integral over the zero modes. For the B-twisted LG theory on a Riemann surface without boundaries, 
the zero modes come from constant scalars 
$z_i, \bar{z_i}, \psi^{\bi}_{\pm}$, and closed 1-forms $\psi_{\pm}^i$. For the disk topology, there are no $\psi^i_{\pm}$ zero modes. 
Furthermore, although the boundary condition does not affect the bosonic zero modes, it leads to the following relation among the 
fermionic zero modes:
$$\psi^{\bi}_- \;=\; \psi^{\bi}_+$$

The path integral now reduces to an ordinary integral with measure
$$\int\prod_{i=1}^n dz_id\bar{z}_i\cdot\prod_{\bi=1}^n d\psi^{\bi}$$
The integrand consists of two factors. The modified bulk action (cf. Section 2) contributes, apart from an unimportant numerical factor, 
$$\exp\big(-\la^2|\part_iW|^2+i\la(W-\bar{W})\big)$$
Although we record explicit $\la$-dependence here, it is understood that we take $\la\to0$ in the end. If the theory were bosonic, 
the presence of a boundary would introduce an extra factor, the holonomy of the connection $A$ along $\part\Sigma$:
$${\rm Tr}\,{\rm Hol}_{\part\Sigma}(A).$$
What we need is a supersymmetric extension which incorporates the the tachyon. Such an extension is well-known and is given by (see for example
\cite{ttu})
$$\str\,P\exp\Big(\la\oint_{\part\Sigma}M(\tau)\,d\tau\Big)$$
with
$$M(\tau) \;=\; \begin{pmatrix}  -T\bar{T} & \psi^{\bi}D_{\bi}T\\ \psi^{\bi}D_{\bi}\bar{T} & -\bar{T}T \end{pmatrix}.$$
In general, $M$ also contains $\dot{z}^I A_I$ and $\psi^I\psi^J F_{IJ}$ in the diagonal entries, but for the zero-mode integration they do not appear. 
The former term is not there since $\dot{z}=0$. The latter term is not there because the curvature is of type $(1,1)$ and must couple to 
$\psi^i\psi^{\bj}$; however, there is no $\psi^i$ zero mode. Note also that the supersymmetric analogue of the Wilson loop
involves a super-trace, rather than the ordinary trace. This arises as a consequence of the GSO projection generalized
to the case of the brane-anti-brane system~\cite{Witten_DK}.

Since we are dealing with zero modes, the path ordering has no effect. 
Recalling that $T=F+\bar{G}$ and taking the trivialization $A_{\bi}=0$, the disk partition function in the limit $\lambda\ra 0$ becomes
$$
Z\;=\;\int\prod_{i} d^2z_id\psi^{\bi}\; 
\str\l[\exp\,\la\big(\psi^{\bi}\part_{\bi}D^\dag - DD^\dagger - D^\dagger D\big)\r]
$$

To evaluate this finite-dimensional integral, we will use the same trick as that employed by Vafa in the closed string case~\cite{Vafa}.
Namely, we will first consider the situation where all critical points of $W$ are isolated and non-degenerate. As explained
in Ref.~\cite{us}, in this case topological correlators are sums over all critical points, and each term in the sum can be evaluated
by expanding everything in Taylor series to a first non-trivial order. In particular, the superpotential can be approximated by
a quadratic function, and we expect that B-branes can be ``approximated'' by Clifford modules. Below we will show very explicitly how
Clifford modules arise. Once we have an explicit expression for the integral, we show that it can be written as a generalized residue
of a meromorphic $n$-form on $\CC^n$, namely, as Eq.~(\ref{eq:form}). Then we can infer Eq.~(\ref{eq:form}) in the general case
by appealing to the holomorphic dependence of topological correlators on the coefficients of $W$.

So let us choose coordinates $z^i$ centered on the critical point, and let $W=Q_{ij}z^iz^j+\ldots$. Here
$Q\in{\rm Sym}^2(V^*)$ is a non-degenerate symmetric bi-linear form, and dots denote terms of higher order in $z$.
The twisted differential $D$ must vanish at $z=0$ for the contribution to the integral to be non-zero (see above), so
its Taylor series expansion starts as follows:
$$
D \;=\; z^i\Ga_i+\ldots .
$$
Here $\Gamma^i$ are $z$-independent odd operators on the graded vector space $M$, where $M$ is defined by the condition
that the graded holomorphic vector bundle $E_1\op E_2$ be $M\otimes_{\CC}\O_X$. The equation $D^2=W$ implies
that $\Ga_i$ satisfy the Clifford algebra relations
$$
\Ga_i\Ga_j+\Ga_j\Ga_i\;=\;2Q_{ij}.
$$  
Thus $M$ is a graded Clifford module, as anticipated. 

All graded Clifford modules are isomorphic to modules
of the form $S\ot_\CC V$, where $S$ is an irreducible Clifford module (spinor module, see Appendix of Ref.~\cite{us}), and 
$V$ is a vector space. Actually, there is a small difference between even and odd $n$: for $n$ even there are two inequivalent
spinor modules which differ by a parity reversal, while for $n$ odd there is only one spinor module, since the parity reversal can
be undone by an automorphism. This means that for $n$ even $V$ should be a graded vector space, while for $n$ odd $V$ is not graded.

We replace $\phi$ by its value at $z=0$ which we denote $\phi(0)$. Since $\phi$ supercommutes with $D$, $\phi(0)$
must supercommute with all $\Ga_i$. In the case when $n$ is even, this implies that $\phi(0)\in \End(V)\ot \End(S)$ has 
the form $a\ot 1$, where $a\in \End(V)$. In the case when $n$ is odd, we have two possibilities for $\phi(0)$: either 
$\phi(0)=a\ot 1$, or $\phi(0)=a\ot (-1)^F \Ga_1\ldots \Ga_n.$ Here $(-1)^F$ denotes the grading operator on $S$, and $a$ is an
arbitrary element of $\End(V)$. 

To simplify the integral still further, note that although superficially it depends both on the (flat) K\"ahler metric on $\CC^n$
and a Hermitian metric on the Clifford module $M$, in fact topological correlators cannot depend on either. This is a general
property of topological correlators in the B-model, but we can make it explicit in the case at hand. We can rewrite the integral 
over bosonic and fermionic variables as an integral of an inhomogeneous differential form:
$$
Z=\;\int_{\CC^n} \str\left[ e^{\lambda [\bpartial-D,D^\dagger]}\alpha(0)\phi(0)\right] dz_1\ldots dz_n.
$$
Hermitian metric enters only through $D^\dag$, and the latter
appears only in the anti-commutator with $\bpartial-D$. Now recall that the latter combination is the total (bulk plus boundary)
BRST charge in the zero-mode approximation. A standard argument then implies that the integral
does not depend on the choice of Hermitian metrics. We can use this freedom to simplify the integral further. First, we
can use a linear change of coordinates $z_i$ to bring the matrix $Q$ to the standard form (identity matrix). The effect of this is to
multiply the integral by $H^{-1/2}$, where $H=\det Q$. Second, we can choose the metric on the vector space $M$ so that $\Ga_i^\dag=\Ga_i$. 
The integral thus becomes
$$
Z=H^{-1/2}\;\int_\CC \str\left[e^{\lambda d\bz^i\Gamma_i-2\lambda z^i\bz^i}\alpha(0)\phi(0)\right]dz_1\ldots dz_n.
$$
Up to an unessential numerical factor, this is equal to
$$
Z=\frac{1}{n!} \alpha(0)H^{-1/2}\;\str\left(\Ga_1\ldots\Ga_n\phi(0)\right).
$$
Note that this is independent of $\lambda$, as expected on general grounds.
Returning to the original coordinates, we get:
$$
Z=\frac{1}{n!}\alpha(0)H^{-1}\;\str\left(\Ga_1\ldots\Ga_n \phi(0)\right).
$$
This is a contribution of a single isolated nondegenerate critical point. If there are several such points $p_1,\ldots,p_k$,
each contributes independently:
$$
Z=\frac{1}{n!}\sum_i \alpha(p_i)H(p_i)^{-1}\; \str\left(\Ga_1\ldots\Ga_n\phi(p_i)\right).
$$
Now recall that one of the properties of the generalized residue is the following identity, holding whenever the superpotential
has only non-degenerate critical points:
$$
\frac{1}{(2\pi i)^n}\oint \frac{f(z)}{\partial_1 W\ldots \partial_n W}=\sum_i H^{-1}(p_i) f(p_i).
$$
Here $f(z)$ is an arbitrary holomorphic function on $\CC^n$. Since $\Gamma_i=\partial_i D$, it follows that $Z$ is given by
Eq.~(\ref{eq:form}).

Suppose now that $W$ has degenerate critical point(s). Consider the following deformation
$$W(z)\;\to\; W(z) + \ep\,h(z), \qquad \ep\in\CC$$
where $h(z)$ is a holomorphic function such that the deformed superpotential has only non-degenerate critical points and the same asymptotic 
behavior as $W(z)$ (i.e. $h(z)$ does not bring in extra critical points from the infinity). 
In the limit $\ep\to0$ one recovers the original theory as some or all non-degenerate critical points of the deformed theory merge. 
It is quite reasonable to expect that all B-branes associated with a degenerate critical point derive from B-branes associated with the 
merging simple critical points for certain choices of $h(z)$. By holomorphy, the disk correlators should still be given by (\ref{eq:form}).
This concludes the derivation of Eq.~(\ref{eq:form}).

It is instructive to check that Eq.~(\ref{eq:form}) enjoys the properties expected from a topological disk correlator. We first show that 
the formula (\ref{eq:form}) leads to the  correct selection rule for the disk correlators. Recall that a generic superpotential leaves unbroken 
a $\ZZ_2$ R-symmetry, which gives the $\ZZ_2$-grading of the boundary operators. For odd $n$ this symmetry is anomalous on a disk, thus
the disk correlator vanishes unless $$ \deg(\phi) + n = 0 \mod 2$$
It is easy to see that the formula (\ref{eq:form}) obeys this selection rule. 

We proceed to show that (\ref{eq:form}) vanishes whenever $\alpha$ or $\phi$ is BRST-exact. For the bulk insertion $\alpha$, 
this means that the residue must vanish if $\alpha$ belongs to the ideal generated by $\partial W$. This follows from the basic
property of the residue: 
$$
\oint \frac{f(z)}{\partial_1 W\ldots \partial_n W}
$$ 
vanishes whenever $f$ belongs to the above-mentioned ideal. Now let us assume that $\phi$ is BRST exact, i.e. $\phi=[D,a]$, 
with $a\in\End E$ satisfying $\deg\,a+n=1\mod2$. Note that the twisted differential satisfies 
$$\part D\cdot D + D\cdot\part D \;=\; \part W.$$
Since an insertion of $\part W$ makes the residue vanish, one can effectively treat $\part D$ and $D$ as anti-commuting. 
It is then an easy computation to show that the disk correlator given by (\ref{eq:form}) for $\phi = [D,a]$ indeed vanishes.

We make one more consistency check before closing this section. In general there is an ordering issue if multiple boundary operators 
are inserted, and the correlator must preserve a super-cyclic symmetry. This is already visible with two boundary operators 
$a,b$. Suppose $a$ and $b$ are initially close to each other. By moving $a$ around $\part\Sigma$ to the other side of $b$, the 
disk correlator must flip sign if both $a$ and $b$ have odd degrees, and must be unchanged otherwise. Furthermore, if one interprets 
$\part D$'s appearing in (\ref{eq:form}) as ``operators'' inserted on the boundary,\footnote{They are not honest boundary operators since 
they do not live in $\D$-cohomology} they should super-commute with true boundary operators for consistency. In the following we 
show that (\ref{eq:form}) does satisfy these consistency requirements.

First, let $a$ be a boundary operator, then $[D,a]=0$. Acting by $\part$, one gets
$$[\part D,a] = - [D, \part a]$$
Since $\part_ia$ are holomorphic sections of $\End E$, the right hand side is a BRST-exact operator. 
Therefore $\part D$ and $a$ can be regarded as super-commuting, just as anticipated.

Now assume that $a$ and $b$ are two BRST-invariant boundary operators. We want to show that the disk correlator 
for $\phi=[a,b]$ as given by (\ref{eq:form}) vanishes. 
The proof is straightforward. As always, it suffices to consider the case when
$$\deg\,a+\deg\,b + n =0\mod2.$$
We already showed that one can make the following substitution
$$\str\big[(\part D)^n a\cdot b\big] \ra 
(-1)^{n\cdot\deg\,a}\str\big[a\,(\part D)^n  b\big].$$
Using the identity $(-1)^Fa(-1)^F = (-1)^{\deg\,a}a$, one immediately gets 
$$\big\langle\alpha_{\rm bulk}\cdot [a,b]\,\big\rangle_{\rm disk} \;=\; 0.$$
This establishes the fact that (\ref{eq:form}) posesses the super-cyclic symmetry, as expected on general grounds.

\subsection{Correlators on arbitrary Riemann surfaces with boundaries}

Let us now consider the case of a general oriented Riemann surface $\Sigma$, which has $g\geq0$ handles and $h\geq1$ boundary components
$C_1,C_2,\ldots,C_h$ each of which is diffeomorphic to a circle. The orientation of $\Sigma$ induces a natural orientation on the $C$'s. 
Without loss of generality we may consider the case when there is a single bulk operator insertion $\alpha$, and a single boundary operator
insertion $\phi_i$ at every boundary component $C_i$. 

We first focus on the case $h=1$, i.e. a disk with several handles attached. One could in principle evaluate the path integral explicitly, 
as is done above, but there is a far more efficient way to arrive at the answer. We note that this correlator can be thought of
as an inner product of two closed string states, using the usual closed-string topological metric. 
The first closed string state is obtained by performing a path integral over a
two-dimensional world-sheet which looks like a genus-$g$ oriented surface with a circular hole, and an insertion of $\alpha$.
We will denote this state by $\vert \alpha,g\rangle.$ From the results of Ref.~\cite{Vafa} it follows easily that the corresponding 
element in $\CC[z_1,\ldots,z_n]/\partial W$ is represented by
$$
\vert \alpha,g\rangle=\alpha(z) H(z)^g ,
$$
where $H(z)$ is the Hessian of $W(z)$. The second closed string state is the boundary state of the B-brane with an insertion of $\phi$
on the boundary.\footnote{Here we use the term ``boundary state'' somewhat unconventionally. The usual boundary state is obtained
by setting $\phi=1$.} From our result for the $g=0,h=1$ correlator we infer that this boundary state is
$$
\vert D,\phi\rangle=\frac{1}{n!}\str \left((\partial D(z))^{\wedge n}\phi(z)\right).
$$
Here it is understood that the overall factor $dz_1\wedge\ldots \wedge dz_n$ is removed, so that the boundary state
is a holomorphic function, rather than a holomorphic $n$-form.
Hence the correlator of interest is given by
$$
\langle \alpha,g\vert D,\phi\rangle=\;\frac1{(2\pi i)^n n!} \oint \frac{\alpha H^g \cdot 
\str\big[(\partial D)^{\wedge n}\phi\big]}{\part_1W\part_2W\ldots\part_nW}
$$

If we have more than one boundary component, we can follow a similar strategy. We take a genus-$g$ closed
string amplitude with $h+1$ operator insertions, and replace $h$ of them with the boundary states corresponding
to $h$ boundary circles $C_i$. We will denote by $D_i$ the boundary part of the BRST operator for the $i^{\rm th}$ brane.
It follows that for general $g$ and $h$ the topological correlator is given by
$$
\frac{1}{(2\pi i)^n (n!)^h} \oint \frac{\alpha H^g}{\part_1W\part_2W\ldots\part_nW} \cdot \prod_{i=1}^h \str\big[(\partial D_i)^{\wedge n}\phi_i\big].
$$
This generalizes Vafa's formula to the open string case. Note that since we removed the factors $dz_1\wedge\ldots\wedge dz_n$ from the
boundary states, the integrand is a meromorphic function. To compute the correlator, one has to compute its residue, i.e. to multiply
it by $dz_1\wedge\ldots\wedge dz_n$ and integrate over a union of Lagrangian tori given by $|\partial_i W|=\epsilon_i$.

\section{Concluding remarks}

We have computed all correlators in the B-twisted Landau-Ginzburg model with boundaries. Perhaps the most interesting extension of these results
would be a computation of the space-time superpotential
in Type II compactifications on Calabi-Yau manifolds with D-branes. At special points in the moduli space such compactification
are described by Landau-Ginzburg orbifolds (Gepner models), and one could hope that the methods of this paper
could be used to determine the {\it exact} superpotential for both open and closed-sector moduli. To achieve this goal, one
must generalize the formalism to orbifolds. Let us sketch how this could be done. 

Suppose that a finite group $H$ acts on $\CC^n$ while preserving both the complex structure and the superpotential $W$. 
Then we can form a semi-direct product of the CDG algebra $(\O,0,W)$ and the group algebra of $H$. It is natural to expect that
CDG modules over this CDG algebra are topological D-branes in the corresponding Landau-Ginzburg orbifold. One can also think of
these CDG modules as equivariant CDG modules over the original CDG algebra $(\O,0,W)$. Let us explain what this means in concrete terms.
Given a CDG module $M=M_+\op M_-$, we need to specify on $M_+$ and $M_-$ representations $R_+$ and $R_-$ of $H$ such that the twisted differential
$D$ intertwines them:
$$
R_+(h) F=F R_-(h),\quad R_-(h) G= G R_+(h),\quad \forall h\in H.
$$

For example, consider the case $n=1$, $W=z^k$. This LG model is known as the $(k-2)^{\rm th}$ minimal model. As mentioned above, any
topological D-brane in this model is equivalent to a direct sum of several indecomposable D-branes, which correspond to the following factorization of
$W$:
$$
F=z^\ell, \quad G=z^{k-\ell},\quad \ell=1,\ldots,k-1.
$$
The modules $M_+$ and $M_-$ are free modules of rank one for all these branes.
Consider now the group $H=\ZZ_k$ which acts by $z\mapsto \beta z$, where $\beta$ is the primitive $k$-th root of unity.
Any representation of $H$ on $M_\pm$ is specified by a single integer $p$ (weight) which runs from $0$ to $k-1$.
Further, since $F$ has weight $\ell$, $H$-equivariance requires that the weight of $M_-$ be equal to the weight of $M_+$ plus
$\ell$. Thus an indecomposable D-brane in an orbifolded theory is a pair $(M,p)$, where $M$ is an indecomposable D-brane in the
original theory, and $p$ is an integer modulo $k$. Since orbifolding creates $k$ copies of each D-brane, the 
total number of indecomposable D-branes in the
orbifolded model is $k(k-1)$. We can check this result using mirror symmetry. The mirror of this LG orbifold is
the unorbifolded LG model with the same superpotential $W=z^k$, hence we expect that there are $k(k-1)$ A-branes in the latter model.
This is indeed the case~\cite{HIV}. Furthermore, one can check the category of B-branes in the orbifolded LG model is
equivalent to the category of A-branes in the unorbifolded model~\cite{Orlovpriv}. This amounts to a proof of the Homological
Mirror Symmetry Conjecture in a simple (non-geometric) case.

Many string orbifolds (including Gepner models) involve not only geometric action on the target space, but also some world-sheet symmetries.
One such symmetry is $\FL$, where $F_L$ is the left-moving fermion number.
The topological version of $\FL$ is the symmetry which reverses the grading of every CDG module $M$ and exchanges $F$ and $G$.
Orbifolding with respect to this symmetry does the following. First of all, we need to keep only CDG modules which
are invariant with respect to this symmetry. In the above example this means that we keep only sums of
D-branes correponding to $\ell=\ell_0$ and $\ell=k-\ell_0$, and, for even $k$, also a brane with $\ell=k/2$.
Second, an object of the new category should really be a pair $(M,\chi)$, where $M$ is a CDG module, and $\chi$ is
an isomorphism between $M$ and its grade-reversal. In the above example, this has the effect that the self-dual brane
with $\ell=k/2$ actually splits into two inequivalent branes. 

It is an interesting problem to compute general bulk-boundary correlators in LG orbifolds, including bulk twisted-sector
states. We hope to address this problem elsewhere.

\section*{Acknowledgments}

We are grateful to Vladimir Baranovsky and Dmitri Orlov for useful conversations. This work was supported in part 
by the DOE grant DE-FG03-92-ER40701.

\end{document}